\documentclass[aps,showpacs,preprintnumbers,amsmath, amssymb]{revtex4}

\usepackage{float}
\usepackage{graphics,epsfig}
\usepackage{graphicx}
\usepackage{dcolumn}
\usepackage{bm}

\begin{document}
\baselineskip=0.8 cm

\title{{\bf Scalar condensation behaviors around regular Neumann reflecting stars}}
\author{Yan Peng$^{1}$\footnote{yanpengphy@163.com}}
\affiliation{\\$^{1}$ School of Mathematical Sciences, Qufu Normal University, Qufu, Shandong 273165, China}

\vspace*{0.2cm}
\begin{abstract}
\baselineskip=0.6 cm
\begin{center}
{\bf Abstract}
\end{center}

We study static massive scalar field condensations in the
regular asymptotically flat reflecting star background.
We impose Neumann reflecting surface boundary conditions for the scalar field.
We show that the no hair theorem holds in the neutral reflecting star background.
For charged reflecting stars, we provide bounds
for radii of hairy reflecting stars. Below the lower bound,
there is no regular compact reflecting star and a black hole
will form. Above the upper bound, the scalar field
cannot condense around the reflecting star or no hair theorems exist.
And in between the bounds, we obtain scalar configurations
supported by Neumann reflecting stars.

\end{abstract}

\maketitle
\newpage
\vspace*{0.2cm}

\section{Introduction}

The classical black holes describe compact configurations that
irreversibly absorb matter and radiation fields at a critical radius
known as the event horizon.
It was believed that the absorbing horizon leads to the famous no scalar hair theorem
that static scalar fields cannot exist around a asymptotically flat black hole,
for references see \cite{Bekenstein}-\cite{Brihaye} and reviews see \cite{Bekenstein-1,CAER}.
However, Hod recently found that this no hair property is not restricted to spacetime
with a horizon and there is also no scalar hair theorem for
regular neutral Dirichlet reflecting stars in the asymptotically flat background \cite{Hod-3}.
This no hair theorem in the reflecting star background opens up an
interesting topic of deriving no scalar hair behaviors in horizonless gravities.

In fact, this no hair theorem also holds in other regular spacetime.
In the asymptotically flat background,
it was proved that regular neutral reflecting stars cannot support
massless scalar fields nonminimally coupled to the gravity \cite{Hod-4}.
This no hair behavior was also observed in regular asymptotically dS reflecting stars \cite{Bhattacharjee}.
On the other aspects of black holes, it has been proved in \cite{Hod-5,Hod-6} that
Reissner-Nordstr$\ddot{o}$m black holes are stable even under
charged scalar perturbations in accordance with the no scalar hair theorem of black holes \cite{AEB}.
Very differently, scalar fields can condense around a charged reflecting shell
in the limit that the spacetime outside the shell is flat \cite{Hod-7,Hod-8}.
In fact, the existence of scalar configurations doesn't depend on the flat spacetime limit.
In a curved gravity, static scalar fields can also condense
around the charged Dirichlet reflecting star \cite{Hod-9,YP2018}.
An interesting property is that hairy Dirichlet reflecting shell and star
radii are discrete and have bounds \cite{Hod-7,YP2018,YP20183,YP20184}.
One naturally wonders what is the case when imposing Neumann reflecting boundary instead of the above
Dirichlet reflecting boundary.
At present, charged scalar field configurations were constructed in the charged Neumann reflecting star background
on conditions that the star charge and mass is very small compared to the star radius \cite{Hod-9}.
Along this line, it is meaningful to extend discussions in \cite{Hod-9} to construct more general charged hairy Neumann reflecting
stars. In addition, it is interesting to search for bounds of hairy Neumann reflecting star radii.
And it is also interesting to examine whether there is still no hair
theorem for neutral Neumann reflecting stars.

The next sections are planed as follows. In section II,
we show the no hair behavior in the neutral Neumann reflecting star background.
In part A of section III, we provide bounds for
radii of charged hairy Neumann reflecting stars.
And in part B of section III, we construct scalar
field configurations supported by charged Neumann reflecting stars.
The last section is devoted to our main results.

\section{The no hair theorem for neutral Neumann reflecting stars}

We consider the system of a scalar field nonlinearly coupled to a reflecting star
in the four dimensional asymptotically flat gravity.
We define the radial coordinate $r=r_{s}$ as the radius of the reflecting star.
And the corresponding Lagrange density of matter fields is
\begin{eqnarray}\label{lagrange-1}
\mathcal{L}=-|\nabla \psi|^{2}-m^{2}\psi^{2},
\end{eqnarray}
where $m$ is the mass of the scalar field with only radial dependence in the form $\psi=\psi(r)$.

The spherically symmetric star geometry deformed by matter fields
can be expressed as \cite{Chandrasekhar}
\begin{eqnarray}\label{AdSBH}
ds^{2}&=&-e^{\nu}dt^{2}+e^{\lambda}dr^{2}+r^{2}(d\theta^{2}+sin^{2}\theta d\varphi^{2}),
\end{eqnarray}
where $\nu$ and $\lambda$ are functions of r satisfying $\nu(\infty)=\lambda(\infty)=0$.

From above assumptions, we obtain the scalar field equation in the form
\begin{eqnarray}\label{BHg}
\psi''+\frac{1}{2}(\frac{4}{r}+\nu'-\lambda')\psi'-m^{2}e^{\lambda}\psi=0.
\end{eqnarray}

Near the infinity boundary $r\rightarrow\infty$, the scalar field behaves as
\begin{eqnarray}\label{InfBH}
&&\psi\thickapprox A\cdot\frac{1}{r}e^{m r}+B\cdot\frac{1}{r}e^{-m r}+\cdots.
\end{eqnarray}
For massive scalar fields with $m>0$, the second scalar operator $B$ is normalizable
and we fix $A=0$. So the boundary condition at the infinity can be expressed as
\begin{equation}
\psi(\infty)=0.
\end{equation}

And at the surface of the star, we impose Neumann reflecting boundary conditions as
 \begin{equation}
d\psi(r_{s})/dr=0.\\
\end{equation}

Firstly, we show that the nontrivial scalar field
$\psi\not\equiv 0$ cannot have zero points. Otherwise,  if there is a
coordinate $r_{*}$ satisfying $\psi(r_{*})=0$ together with (5),
we must have at least one extremum point $r=r_{peak}$
(positive maximum extremum point or negative minimum extremum point)
between the zero point coordinate $r_{*}$ and the infinity boundary.
At this extremum point, the scalar field is characterized by the relation
\begin{eqnarray}\label{InfBH}
\{ \psi'=0~~~~and~~~~\psi \psi''\leqslant0\}~~~~for~~~~r=r_{peak}.
\end{eqnarray}

It leads to the inequality at this extremum point as
\begin{eqnarray}\label{BHg}
\psi\psi''+\frac{1}{2}(\frac{4}{r}+\nu'-\lambda')\psi\psi'-m^{2}e^{\lambda}\psi\psi<0~~~~for~~~~r=r_{peak}
\end{eqnarray}
in contradiction with the equation (3).
So we arrive at a conclusion that the scalar field has no zero points.
Then there are two cases: $\psi>0$ and $\psi<0$.

According to the above conclusion, we suppose the scalar field to be positive in the following
and the discussion is qualitatively the same for negative scalar fields.
Now, we demonstrate that there is $\psi''(r_{s})\leqslant 0$ with proof by contradiction.
If $\psi''(r_{s})>0$ and also considering the Neumann reflecting condition $\psi'(r_{s})=0$,
we will have $\psi'(r)>0$ around $r_{s}$.
It means the positive scalar field first increases to be more larger and approaches
zero at the infinity. In this case, the scalar field must have one positive maximum extremum point.
At this maximum extremum point, it holds a relation in the form of (8), which is however
in contradict with the equation (3).
So the assumption is invalid and there is $\psi''(r_{s})\leqslant 0$ in cases of $\psi>0$.
We can similarly show that it holds $\psi''(r_{s})\geqslant 0$ for $\psi<0$.
Then, we obtain following relations
\begin{eqnarray}\label{InfBH}
\{ \psi'=0~~~~and~~~~\psi \psi''\leqslant0\}~~~~for~~~~r=r_{s}.
\end{eqnarray}

From (9), we obtain the inequality at the star surface as
\begin{eqnarray}\label{BHg}
\psi\psi''+\frac{1}{2}(\frac{4}{r}+\nu'-\lambda')\psi\psi'-m^{2}e^{\lambda}\psi\psi<0~~~~for~~~~r=r_{s}
\end{eqnarray}
in contradiction with the equation (3). So the only acceptable solution is
the trivial scalar field $\psi\equiv 0$ or the regular neutral
Neumann reflecting star cannot support static massive scalar fields.
It is directly to check that the no hair theorem also holds for more general models with scalar field
potentials $V(\psi^{2})$ satisfying $V'(\cdot)>0$ and the lagrange density (1) corresponds to
the special case of $V'(\cdot)=m^2>0$.

Here we find that there is no scalar hair theorem in the
Neumann reflecting star. In addition, it is known that the no scalar hair behavior also exists
in the Dirichlet reflecting star background \cite{Hod-3}. So it is possible that a similar no hair theorem can also be established
for more general Robin boundary conditions.
And we mention that scalar condensation behaviors with Robin boundary conditions imposed to either the
mirror-like cavity or the AdS boundary were investigated in \cite{HRC}.
We plan to study scalar condensation with Robin boundary conditions
imposed at the regular star surface in the next work.

\section{Scalar condensations around charged Neumann reflecting stars}

\subsection{Bounds for radii of hairy charged Neumann reflecting stars}

In this part, we consider the scalar field coupled to the charged Neumann reflecting star
in the probe limit and the corresponding Lagrange density is given by
\begin{eqnarray}\label{lagrange-1}
\mathcal{L}=-\frac{1}{4}F^{MN}F_{MN}-|\nabla_{\mu} \psi-q A_{\mu}\psi|^{2}-m^{2}\psi^{2},
\end{eqnarray}
where q and $m$ are the charge and mass of the scalar field $\psi(r)$ respectively.
And $A_{\mu}$ stands for the ordinary Maxwell field with only non-zero t component as $A_{\mu}=-\frac{Q}{r}dt$.

The background solution is a charged spherically symmetric reflecting star
in the form \cite{Chandrasekhar}
\begin{eqnarray}\label{AdSBH}
ds^{2}&=&-g(r)dt^{2}+\frac{dr^{2}}{g(r)}+r^{2}(d\theta^{2}+sin^{2}\theta d\varphi^{2})
\end{eqnarray}
where $g(r)=1-\frac{2M}{r}+\frac{Q^2}{r^2}$ with $M$ and $Q$
as the mass and charge of the star respectively.
Since we study the regular spacetime without a horizon, we assume
$r_{s}>M+\sqrt{M^2-Q^2}$ and $M\geqslant Q$.

The scalar field equation of motion is
\begin{eqnarray}\label{BHg}
\psi''+(\frac{2}{r}+\frac{g'}{g})\psi'+(\frac{q^2Q^2}{r^2g^2}-\frac{m^2}{g})\psi=0
\end{eqnarray}
with $g=1-\frac{2M}{r}+\frac{Q^2}{r^2}$.

Defining a radial function $\tilde{\psi}=\sqrt{r}\psi$,
one obtains the new differential equation
\begin{eqnarray}\label{BHg}
r^2\tilde{\psi}''+(r+\frac{r^2g'}{g})\tilde{\psi}'+(-\frac{1}{4}-\frac{rg'}{2g}+\frac{q^2Q^2}{g^2}-\frac{m^2r^2}{g})\tilde{\psi}=0.
\end{eqnarray}

According to (4) and (5), we have the infinity boundary condition
\begin{eqnarray}\label{BHg}
\tilde{\psi}(\infty)=0.
\end{eqnarray}

Now we show that nontrivial $\tilde{\psi}\not\equiv0$ must have one extremum point.
In cases of $\psi(r_{s})=0$, there is $\tilde{\psi}(r_{s})=0$
and one extremum point $r=\tilde{r}_{peak}$ (positive maximum extremum point or negative minimum extremum point)
must exist between star surface $r_{s}$ and the infinity \cite{Hod-3}. In another case of $\psi(r_{s})>0$, we have
$\tilde{\psi}'=(\sqrt{r}\psi)'=\frac{1}{2\sqrt{r}}\psi+\sqrt{r}\psi'=\frac{1}{2\sqrt{r}}\psi>0$ for $r=r_{s}$
and $\tilde{\psi}$ increases as a function of r around $r=r_{s}$.
Then $\tilde{\psi}$ must have one positive maximum extremum point $r=\tilde{r}_{peak}$ according to
the fact that $\tilde{\psi}$ firstly increases to be more positive
and approaches zero at the infinity.
And for other cases of $\psi(r_{s})<0$, we can similarly obtain the negative minimum extremum point $r=\tilde{r}_{peak}$.
At these extremum points, the radial function $\tilde{\psi}$ is characterized by
\begin{eqnarray}\label{InfBH}
\{ \tilde{\psi}'=0~~~~and~~~~\tilde{\psi} \tilde{\psi}''\leqslant0\}~~~~for~~~~r=\tilde{r}_{peak}.
\end{eqnarray}

According to the relations (14) and (16), we arrive at the inequality
\begin{eqnarray}\label{BHg}
-\frac{1}{4}-\frac{rg'}{2g}+\frac{q^2Q^2}{g^2}-\frac{m^2r^2}{g}\geqslant0~~~for~~~r=\tilde{r}_{peak}.
\end{eqnarray}

Then we have
\begin{eqnarray}\label{BHg}
m^2r^2g\leqslant q^2Q^2-\frac{rgg'}{2}-\frac{1}{4}g^2~~~for~~~r=\tilde{r}_{peak}.
\end{eqnarray}

With assumed regular conditions $r\geqslant r_{s}> M+\sqrt{M^2-Q^2}\geqslant M \geqslant Q$, we have
\begin{eqnarray}\label{BHg}
g=1-\frac{2M}{r}+\frac{Q^2}{r^2}=\frac{1}{r^2}(r^2-2Mr+Q^2)=\frac{1}{r^2}[(r-M)^2-(M^2-Q^2)]\geqslant0,
\end{eqnarray}
\begin{eqnarray}\label{BHg}
rg'=r(1-\frac{2M}{r}+\frac{Q^2}{r^2})'=r(\frac{2M}{r^2}-\frac{2Q^2}{r^3})=\frac{2M}{r}(1-\frac{Q}{r}\frac{Q}{M})\geqslant 0
\end{eqnarray}
and
\begin{eqnarray}\label{BHg}
(r^2g)'=(r^2-2Mr+Q^2)'=2(r-M)\geqslant 0.
\end{eqnarray}

We arrive at
\begin{eqnarray}\label{BHg}
m^2r_{s}^2g(r_{s})\leqslant m^2r^2g(r)\leqslant q^2Q^2-\frac{rgg'}{2}-\frac{1}{4}g^2\leqslant q^2Q^2~~~for~~~r=\tilde{r}_{peak}.
\end{eqnarray}

According to (22), there is
\begin{eqnarray}\label{BHg}
m^2r_{s}^2g(r_{s})\leqslant  q^2Q^2.
\end{eqnarray}

Taking cognizance of the metric solutions, (23) can also be expressed as
\begin{eqnarray}\label{BHg}
m^2r_{s}^2(1-\frac{2M}{r_{s}}+\frac{Q^2}{r_{s}^2})\leqslant  q^2Q^2.
\end{eqnarray}

We can transfer (24) into the form
\begin{eqnarray}\label{BHg}
(m r_{s})^2-(2m M)(m r_{s})+Q^2(m^2-q^2)\leqslant 0.
\end{eqnarray}

Then, we obtain bounds for the radius of the scalar hairy compact reflecting star as
\begin{eqnarray}\label{BHg}
m M+\sqrt{m^{2}(M^2-Q^2)} < m r_{s}\leqslant m M+\sqrt{m^{2}(M^2-Q^2)+q^2Q^2}.
\end{eqnarray}
The lower bound comes from the assumption that
the spacetime is regular or $r_{s}>M+\sqrt{M^2-Q^2}$
and the upper bound can be obtained from (25).
For models with neutral scalar field $q=0$ and other cases in
neutral reflecting stars with $Q=0$, (26) shows that
the upper bound is behind a horizon or no hair theorem holds.
So it is the coupling $qQ$ makes the upper bound larger than the horizon critical points
and then the scalar hair can possibly form in the regular Neumann reflecting star spacetime.

\subsection{Scalar configurations supported by charged Neumann reflecting stars}

In this part, we will extend discussions in \cite{Hod-9} to construct more general hairy charged
Neumann reflecting star solutions by relaxing the condition
that the star charge and mass is very small compared to star radii.
We can simply set $m=1$ in the following with the symmetry of the equation (3) in the form
\begin{eqnarray}\label{BHg}
r\rightarrow k r,~~~~ m\rightarrow m/k,~~~~ M\rightarrow k M,~~~~ Q\rightarrow k Q,~~~~ q\rightarrow q/k.
\end{eqnarray}
Since we impose Neumann reflecting boundary conditions for the scalar field
at the star surface, the star surface can be putted at the extremum point
of the scalar field. In numerical calculations, we will not
expand the scalar field at the star surface. Instead, we fix the scalar field
to be zero at various coordinate $r_{0}$ and integrate the equation from various
$r_{0}$ to the infinity to search for the proper zero point $r_{0}$
satisfying the infinity boundary conditions (4) and (5).
Then, the Neumann reflecting surface $r_{s}$ can be imposed at
extremum points of the obtained scalar field.

We plot the scalar field with $q=2$, $Q=4$ and $M=5$ in the left panel of Fig. 1.
According to results of \cite{YP2018}, we can fix the zero point of the scalar field at $r_{0}\thickapprox11.38808027313$
and the corresponding scalar field asymptotically approaches zero at the infinity.
Integrating the equation from $r_{0}=11.38808027313$ to smaller and larger radial
coordinates in the right panel of Fig. 1, we can obtain various
discrete radii of hairy reflecting stars
at discrete points around $r_{s}\thickapprox12.440,~~10.743,~~9.880, \cdots,~~8.003,\cdots$
between bounds $8< r_{s}\leqslant 5+\sqrt{73}\thickapprox13.544$ according to (26).
Here,  we find that hairy Neumann reflecting star radii are discrete
similar to cases in other regular reflecting object backgrounds \cite{Hod-7,YP2018,YP20183,YP20184,Hod-9}.

\begin{figure}[h]
\includegraphics[width=165pt]{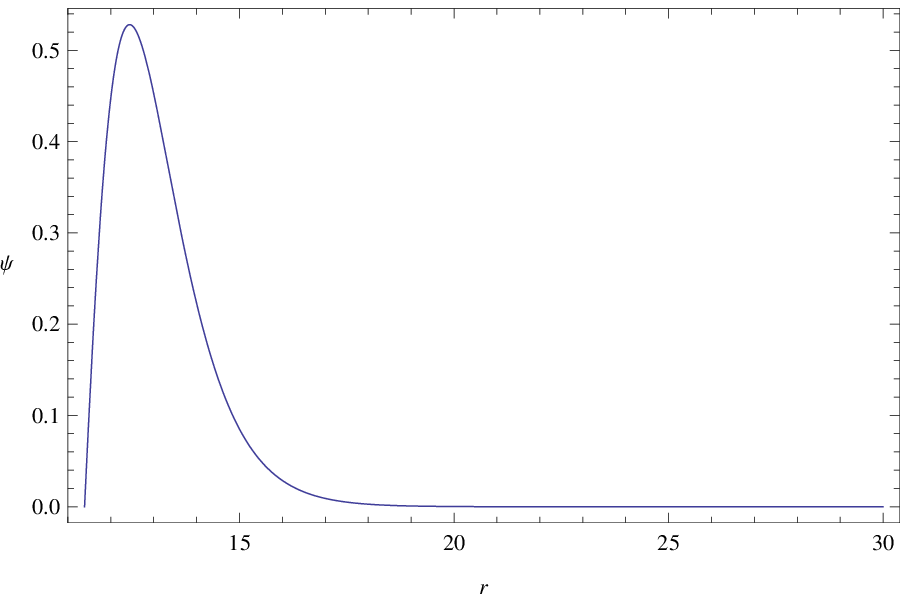}\
\includegraphics[width=168pt]{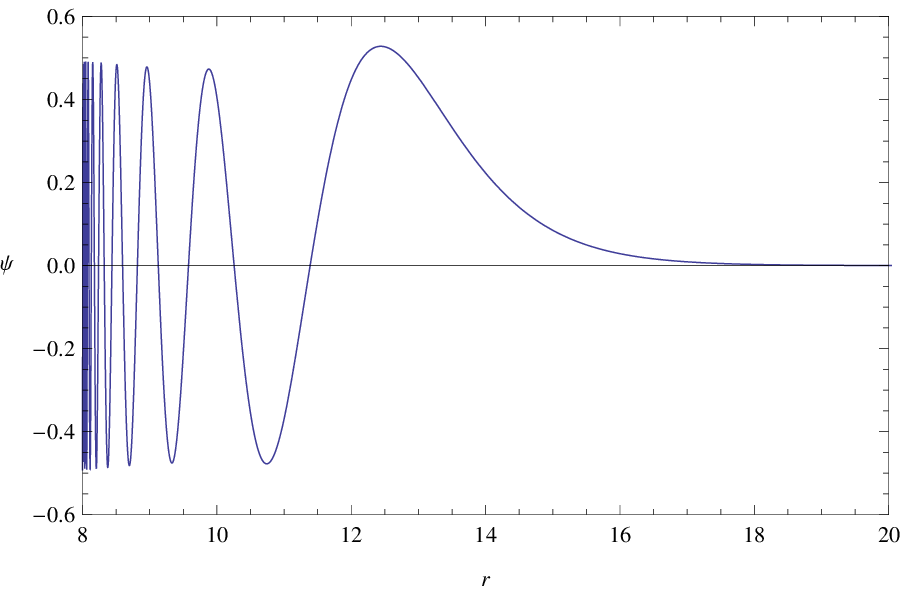}\
\caption{\label{EEntropySoliton} (Color online) We show the scalar field $\psi$
as a function of the radial coordinate r with $q=2$, $Q=4$ and
$M=5$. The left panel is with $r_{0}=11.38808027313$
and the right panel shows various discrete hairy Neumann reflecting star radii.}
\end{figure}

\section{Conclusions}

We studied condensation behaviors of static scalar fields around a Neumann
reflecting star in the asymptotically flat spacetime.
We showed that no scalar hair theorem exists in the neutral Neumann reflecting star background.
For charged Neumann reflecting stars, we
provided bounds for the radius of the scalar hairy star
as $m M+\sqrt{m^{2}(M^2-Q^2)} < m r_{s}\leqslant m M+\sqrt{m^{2}(M^2-Q^2)+q^2Q^2}$,
where m is the scalar field mass, q is the scalar field charge, M is the star mass and Q corresponds to the star charge.
The lower bound is equal to the horizon of a black hole with the same background charge and mass
and there is no regular gravity system below the lower bound.
Above the upper bound, the scalar field cannot condense around charged Neumann reflecting stars
or no scalar hair theorem exists. And in between the bounds, we obtained scalar field configurations
supported by charged Neumann reflecting stars with numerical methods and the hairy star radii are discrete.

\begin{acknowledgments}

We would like to thank the anonymous referee for the constructive suggestions to improve the manuscript.
This work was supported by Shandong Provincial Natural Science Foundation of China under Grant No. ZR2018QA008.

\end{acknowledgments}

\end{document}